# Application of CellDesigner to the Selection of Anticancer Drug Targets: Test Case using P53.

Raúl Isea[1*], Johan Hoebeke[1], Rafael Mayo[2], Fernando Alvarez[3]
and David S. Holmes[4*]

[1]Fundación Instituto de Estudios Avanzados, Valle de Sartenejas, Baruta 1080, Venezuela;
[2]CIEMAT, Av. Complutense de Madrid 22, Spain;
[3]Sección Biomatemática, Facultad de Ciencias, Montevideo 11400, Uruguay;
[4]Center for Bioinformatics and Genome Biology, Fundación Ciencia para la Vida and Facultad de Ciencias Biologicas, Universidad Andrés Bello, Santiago, Chile.

(*) Corresponding author. DSH (dsholmes2000@yahoo.com), RI (raul.isea@gmail.com)

**ABSTRACT**

Cancer is a disease involving many genes, consequently it has been difficult to design anticancer drugs that are efficacious over a broad range of cancers. The robustness of cellular responses to gene knockout and the need to reduce undesirable side effects also contribute to the problem of effective anti-cancer drug design. To promote the successful selection of drug targets, each potential target should be subjected to a systems biology scrutiny to locate effective and specific targets while minimizing undesirable side effects. The gene p53 is considered a good candidate for such a target because it has been implicated in 50% of all cancers and is considered to encode a hub protein that is highly connected to other proteins. Using P53 as a test case, this paper explores the capacity of the systems biology tool, CellDesigner, to aid in the selection of anticancer drug targets and to serve as a teaching resource for human resource development.

## 1. INTRODUCTION

Over the past decade, there has been a significant decrease in the rate that new anticancer drug candidates are being translated into effective therapies in the clinic. In particular, there has been a worrying rise in late-stage attrition in phase 2 and phase 3 [1]. Currently, the two single most important reasons for this attrition, each of which account for 30% of failures are (i) lack of efficacy and (ii) clinical safety or toxicology. These late-stage attrition rates are at the heart of much of the relative decline in productivity of the pharmaceutical industry resulting in significant financial setbacks. In addition, because many patents on the current generation of marketed drugs will expire from 2010 onward, pharmaceutical companies will face the first fall in revenue in four decades.

With advances in human genome biology and the advent of personalized medicine, new opportunities for the discovery of drug targets are arising. The fundamental challenge of anticancer therapy is the need for agents that eliminate cancer cells with a therapeutic index that is safely tolerated by the patient. However, the selection of suitable anticancer drug targets is complicated by the fact that over 100 genes have been shown to be involved in the disease, rendering it necessary to develop multiple therapies. In addition, the cell shows a remarkable degree of robustness, in which impairment of the functions of some genes by mutation or drug targeting can be rescued by the action of other genes. For example, using genetic engineering techniques to knock out the function of over 40% of the approximately 6000 yeast genes one-by-one has little effect on cell growth in rich medium [2, 3].

The transcription factor P53 is encoded by a tumor suppressor gene that can arrest the cell cycle and facilitate apoptosis. P53 functions as a sensor of upstream signals that reflect DNA-damage and cellular stresses such as hypoxia, providing protection against tumor growth, and the action of oncogenes (Myc, Ras, E1A, ß-catenin) [4, 5]. These signals activate latent p53 by enhancing its DNA-binding activity and increasing its stability, resulting in increased levels of 10-100x of the protein. Since enhanced levels of p53 lead to cell cycle-arrest and apoptosis, it is of critical importance that normal cells keep their p53 levels low. This can be accomplished by destabilization of P53 by interaction with the protein Mdm2. Interaction with Mdm2 causes export of P53 from the nucleus to the cytoplasm where it is targeted for proteosomal degradation by interaction with a P53-specific E3 ubiquitin [6]. The gene encoding Mdm2 is activated by P53 but at a time that is relatively late in the cell cycle providing a window of time where P53 can function.

P53 has been considered a suitable target for anticancer therapies because of its widespread implication in about 50% of all cases of cancer [7-9]. It is also a node protein with extensive interaction partners (a hub) and high turnover rate [10]. As such, its function may not easily be replaced by other proteins (knockout inviable) [10]. Several strategies have been explored in attempts to stabilize P53 by preventing or diminishing its interaction with Mdm2 [11-18]. The long term objective is to improve the efficacy of P53 to promote cell apoptosis in cancer cells.

Given the promise of this approach and the substantial body of mathematical modeling and experimental evidence describing the cellular oscillations of the P53-Mdm2 pair, we decided to explore the use of CellDesigner, as a tool for evaluating the network consequences of eliminating the interaction of Mdm2 with P53. CellDesigner can visualize, model and simulate biochemical networks and can integrate with existing databases such as KEGG, PubMed, BioModels [19].

## 2. METHODS

2.1 Systems parameters

The parameters that describe the initial states and oscillatory nature of the p53-Mdm2 interaction in the normal cell were obtained by combining information from several sources [20-23]. These parameters allow the partial differential equations that govern the system to be derived:

$$\frac{\partial x}{\partial t} \equiv \dot{x} = \beta_x \xi - \alpha_x x - \alpha_k y \frac{x}{x+k}$$

[Equation 1]

$$\dot{y} = \alpha_o y_o - \alpha_y y$$

[Equation 2]

$$\dot{y}_o = \beta_y x \bar{\xi} - \alpha_o y_o$$

[Equation 3]

Equations 1 describes the negative feedback loop (defined as x) for P53 in which its activity decreases with time according to the mass-action binding of Mdm2 to P53 resulting in P53 ubiquitination and subsequent proteasomal degradation. Equation 2 describes the transcriptional activity of p53 (defined as y), depending on an initial value and a production rate that decreases with time as Mdm2 activity increases. Equation 3 describes the transcriptional activity of Mdm2 ($y_o$) which decreases from an initial value because of the negative feedback loop imposed by parameters described in the second equation.

For these equations, the variables are derived from [21] the initial values of the variables are: x (0.20), y (0.10), and $y_o$ (0.20). The other parameters are: $\alpha_k$ is the rate of Mdm2-independent degradation of P53 and the P53-dependent Mdm2 production rate (0); $\alpha_y$ is the Mdm2 degradation rate (0.8); $\beta_x$ is the P53 production rate (0.9); $\beta_y$ is the P53-dependent MDM2 production rate (1.1); $\alpha_o$ is the maturation rate (0); k is the P53 threshold for degradation by MDM2 (0.0001) and $\xi$ is equal to 1. On addition of a drug that completely eliminates the P53-Mdm2 interaction, factors are added to the x and $y_o$ equations, where $\delta$ is the drug concentration (0.17 arbitrary units).

2.2 Pathways and Use of CellDesigner

Pathways describing the interaction of P53 with Mdm2 were uploaded from KEGG [24] into CellDesigner using the intuitive interface comprised of graphical notation, model description and an application integration environment as described [19, 25-27].

## 3. RESULTS

Two network models for the interaction of P53 and Mdm2 were generated by CellDesigner. The first network illustrates the negative feedback loop that models the experimentally observed time-delayed oscillatory nature of both P53 and Mdm2 (Fig. 1A). The second network introduces a drug-induced inhibition of the interaction between P53 and Mdm2 that in turn impedes the ubiquitination of P53 and its subsequent degradation by proteosomes (Fig. 1B). The drug is hypothetical and is assumed to cause significant inhibition of the P53/Mdm2 interaction.

Using parameters described in Methods and the networks shown in Fig. 1, CellDesigner was used to predict fluctuations of P53 and Mdm2 for 400 min in normal cells (Fig. 2A) and for 200 min in cells with the drug application (Fig. 2B). The application successfully predicted the time-delayed, oscillatory profiles of the interacting partners in the normal cell with the characteristic 20 min periodicity of fluctuation (Fig. 2A) and predicted a damping out of both P54 and Mdm4 after the addition of the drug (Fig. 2B) that has been observed in mathematical models [29, 30].

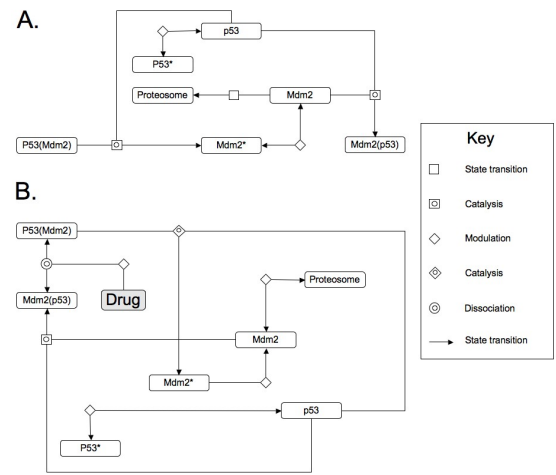

**Figure 1: Model of the interaction of P53 with Mdm2 and the resulting perturbations in the system predicted by CellDesigner in the absence (A) or presence (B) of a hypothetical drug that blocks the interaction between the two proteins. P53\* = P53 precursor, Mdm2\* = Mdm2 precursor, P53(Mdm2) = p53-dependent Mdm2 production rate, Mdm2(P53) = Mdm2-dependen p53 production rate.**

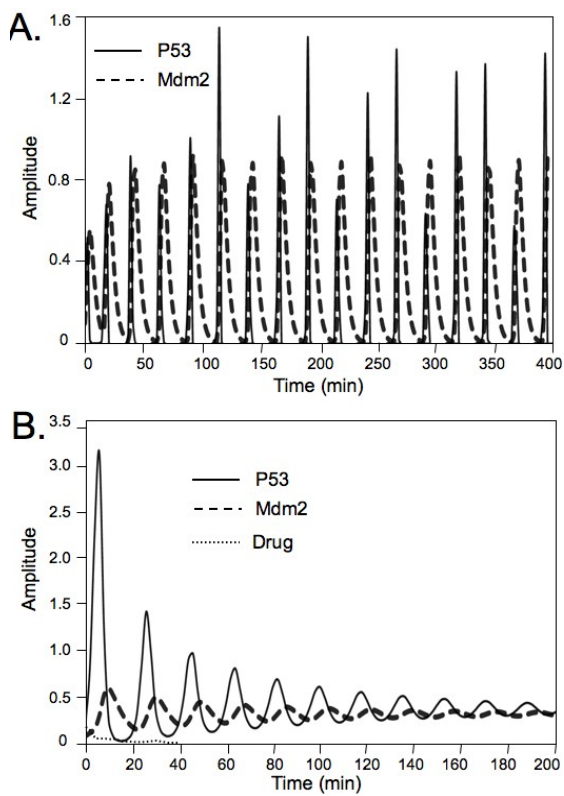

**Figure 2: Prediction of the changes in the amount (amplitude in arbitrary units) of P53 and Mdm2 per cell as a function of time without (A) and with (B) addition of a hypothetical drug that blocks the interaction of the two proteins. The curves were generated using the flow diagram shown in Figure 1.**

## 4. DISCUSSION

The time-delayed, oscillatory behavior of P53 and Mdm2 in the normal cell has been observed experimentally [29] and has been modeled theoretically [28, 30-33]. In addition, mathematical models have been developed that predict damping-out of concentrations of P53 following the perturbation of the interaction of P54 with Mdm2 via a naturally occurring mutation [38] or after treatment with the small molecule inhibitor Nutlin-3 [36]. The problem with these theoretical models is that they are essentially "static". CellDesigner is also able to successfully model the behavior of P54 and Mdm2 in the normal cell (Figs. 1A and 2A) and can also predict damping out when a theoretical drug that prevents P54/Mdm2 interaction is introduced into the system (Figs 1B and 2B). However, in contrast to the previous models, CellDesigner integrates pathway visualization with analysis using Systems Biology Markup Language [38], providing a more dynamic model of cellular processes through the incorporation of gene expression data, subcellular localization information, time-dependent behavior, chemical reactions, kinetic equations etc. a one stop shopping environment. Also, because CellDesigner is open source, custom designed plug-ins can be added when required. CellDesigner has been used to predict suitable drug targets for for methicillin-resistant bacteria [39], but this is the first time that it has been used to examine anti-cancer targets. Because it is open source and manageable in a user-friendly fashion, we have found CellDesigner to be an excellent aid for teaching systems biology at the undergraduate and graduate level.

Other open source services similar to CellDesigner include PathwayStudio [40], Cytoscape [41], VisANT [42], BiologicalNetworks [43] and Teranode Pathway Analytics (available at www.teranode.com/). As increasing abundance of experimental data from high-throughput experimental biology becomes available and as the cost, efficiency and accuracy of this data improves, greater emphasis will be put on systems biology applications such as Celldesigner and others for connecting vast amounts of data and but also capturing usable knowledge in the form of biologically valid relations that research scientists can apply.

The modeling of the P53/Mdm2 interaction network presented here represents a simplistic first step for evaluating anticancer drug targets by CellDesigner. But it should serve as the starting point for several interesting lines of further inquiry including the evaluation of the interactions of P53 with Akt [44], p14ARF and ATM [20], Pten and Pip3 [45] and other members of its interactome. This information needs to be explored in order to provide a more comprehensive understanding of the potential of the P53/Mdm2 interaction as a anticancer drug target.

## 5. ACKNOWLEDMENTS

Work supported by Fondecyt 1090451, DI-UNAB 15-06/I and a Microsoft Sponsored Research Award. Annual Operative Plan of Fundación Instituto de Estudios Avanzados IDEA. This project was carried out under the auspices of the Virtual Institute of Integrative Biology, a shared environment for e-Science.

## 6. REFERENCES

[1] I. Kola and J. Landis, Can the pharmaceutical industry reduce attrition rates?, Nature Reviews Drug Discovery 3 (2004) 711-716.

[2] P. Uetz, L. Giot, G. Cagney, T.A. Mansfield, R.S. Judson, et al., A comprehensive analysis of protein–protein interactions in Saccharomyces cerevisiae, Nature 403 (2000) 623–627.

[3] H. Moriya, Y. Shimizu-Yoshida, and H. Kitano, In vivo robustness analysis of cell division cycle genes in Saccharomyces cerevisiae, PLoS Genet 2 (2006) e111.

[4] W.S El-Deiry, The p53 pathway and cancer therapy, Cancer Journal 11 (1998) 229-236.

[5] D.P. Lane, T.R. Hupp, Drug discovery and p53, Drug Discovery Today 8 (2003) 347-355.

[6] G. Asher, J. Lotem, L. Sachs, C. Kahana, and Y. Shaul, Mdm-2 and ubiquitin-independent p53 proteasomal degradation regulated by NQO1, PNAS 99 (2002) 13125-13130.

[7] H. Jeong, S.P. Mason, A.L. Barabasi, and Z.N. Oltvai, Lethality and centrality in protein networks, Nature 411 (2001) 41–42.

[8] J.B. Pereira-Leal, B. Audit, J.M. Peregrin-Alvarez, C.A. Ouzounis, An exponential core in the heart of the yeast protein-interaction network, Mol Biol Evol 22 (2005) 421–425.

[9] N.N. Batada, L.D. Hurst, M Tyers Evolutionary and physiological importance of hub proteins, PLoS Comput Biol 2 (2006) e88.

[10] Y. Liu and M. Kulesz-Martin, p53 protein at the hub of cellular DNA damage response pathways through sequence-specific and non-sequence-specific DNA binding, Carcinogenesis 22 (2001) 851-860.

[11] L.T. Vassilev, B.T. Vu, B. Graves, et al. In vivo activation of the p53 pathway by small-molecule antagonists of MDM2. Science 303 844-848 (2004).

[12] S. Shangary and S. Wang, Small-Molecule Inhibitors of the MDM2-p53 Protein-Protein Interaction to Reactivate p53 Function: A Novel Approach for Cancer Therapy, Annual review of pharmacology and toxicology, 49 (2009) 223-241.

[13] P. Chène, Inhibition of the p53-MDM2 Interaction: Targeting a Protein-Protein Interface, Mol Cancer Res 2 (2004) 20

[14] C.F. Cheok and D.P. Lane, New developments in small molecules targeting p53 pathways in anticancer therapy, Drug Development Research 69 (2008) 289 – 296.

[15] W. Wang, W.S. El-Deiry, Targeting p53 by PTD-mediated transduction, Trends in Biotechnology 22 (2004) 431-434.

[16] A.V. Gudkov, Therapeutic Strategies Based on Pharmacological Modulation of p53 Pathway, 2 (2005) 225-242.

[17] C. Klein and L.T. Vassilev, Targeting the p53–MDM2 interaction to treat cancer, British Journal of Cancer 91 (2004) 1415–1419.

[18] H.R. Lawrence, Z. Li, M.L. Yip, S.S. Sung, N.J. Lawrence, M.L. McLaughlin, G.J. McManus, M.J. Zaworotko, S.M. Sebti, J. Chen, W.C. Guida, Identification of a disruptor of the MDM2-p53 protein-protein interaction facilitated by high-throughput in silico docking, Bioorg Med Chem Lett. 19 (2009) 3756-3759.

[19] A. Funahashi, A. Jouraku, Y. Matsuoka and H. Kitano, Integration of CellDesigner and SABIO-RK, In Silico Biology 7 (2007) 10.

[20] C.J. Proctor and D.A Gray, Explaining oscillations and variability in the p53-Mdm2 system, BMC Systems Biology 2 (2008) 75.

[21] N. Geva-Zatorsky, N. Rosenfeld, S. Itzkovitz, R. Milo, A. Sigal, E. Dekel, T. Yarnitzky, Y. Liron, P. Polak, G. Lahav, U. Alon, Oscillations and variability in the p53 system. Molecular Systems Biology, 2 (2006) 33.

[22] R.L. Bar-Or, R. Maya, L.A. Segel, U. Alon, A.J. Levine, and M. Oren, Generation of oscillations by the p53-Mdm2 feedback loop: A theoretical and experimental study, PNAS 97 (2000) 11250-11255.

[23] G. Lahav, Oscillations by the p53-Mdm2 feedback loop, Adv Exp Med Biol., 641 (2008) 28-38.

[24] M. Kanehisa, S. Goto, S. Kawashima, Y. Okuno, and M. Hattori, The KEGG Resource for Deciphering the Genome, Nucleic Acids Research 32 (2004) D277-D280.

[25] A. Funahashi, N. Tanimura, M. Morohashi, and H. Kitano, CellDesigner: a process diagram editor for gene-regulatory and biochemical networks, BIOSILICO 1 (2003) 159-162.

[26] H. Kitano, A. Funahashi, Y. Matsuoka, and K. Oda, Using process diagrams for the graphical representation of biological networks, Nat. Biotechnol. 23 (2005) 961-966.

[27] A. Funahashi, Y. Matsuoka, A. Jouraku, M. Morohashi, N. Kikuchi, H. Kitano, CellDesigner 3.5: A Versatile Modeling Tool for Biochemical Networks, Proceedings of the IEEE 96 (2008) 1254 – 1265.

[28] X. Cai, and Z.M. Yuan, Stochastic modeling and simulation of the p53-MDM2/MDMX loop, J. Comput. Biol. 16 (2009) 917-933.

[29] D.A. Hamstra, M.S. Bhojani, L.B. Griffin, B. Laxman, B.D. Ross, A. Rehemtulla, Real-time evaluation of p53 oscillatory behavior in vivo using bioluminescent imaging, Cancer Res. 66 (2006) 7482-7489.

[30] W. Abou-Jaoudé, D.A. Ouattara, M. Kaufman, From structure to dynamics: frequency tuning in the p53-Mdm2 network I. Logical approach, J. Theor. Biol. 258 (2009) 561-577.

[31] K. Puszyński, B. Hat, and T. Lipniacki, Oscillations and bistability in the stochastic model of p53 regulation, J. Theor. Biol. 254 (2008) 452-465.

[32] E. Batchelor, C. S. Mock, I. Bhan, A. Loewer, G. Lahav, Recurrent initiation: a mechanism for triggering p53 pulses in response to DNA damage, Mol Cell. 30 (2008) 277-289.

[33] A. Ciliberto, B. Novak, and J.J. Tyson, Steady states and oscillations in the p53/Mdm2 network. Cell Cycle, 4 (2005) 488-493.

[34] L. Ma, J. Wagner J, J.J. Rice, W. Hu, A.J. Levine, and G.A. Stolovitzky, A plausible model for the digital response of p53 to DNA damage, Proc Natl Acad Sci U S A. 102 (2005)14266-14271.

[35] W. Hu, Z. Feng, L. Ma, J. Wagner, J.J. Rice, G. Stolovitzky, and A.J. Levine, A single nucleotide polymorphism in the MDM2 gene disrupts the oscillation of p53 and MDM2 levels in cells, Cancer Res. 67(2007) 2757-2765.

[36] L.T. Vassilev, B.T. Vu, B. Graves, et al., In vivo activation of the p53 pathway by small-molecule antagonists of MDM2, Science 303 (2004) 844-848.

[37] S. Shangary, and S. Wang, Small-Molecule Inhibitors of the MDM2-p53 Protein-Protein Interaction to Reactivate p53 Function: A Novel Approach for Cancer Therapy, Annu Rev Pharmacol Toxicol. 49 (2009) 223-241.

[38] M. Hucka, A. Finney, H.M. Sauro, H. Bolouri, J.C. Doyle, H. Kitano, et al., The Systems Biology Markup Language (SBML): A Medium for Representation and Exchange of Biochemical Network Models. Bioinformatics, 9(4):524–531, 2003.

[39] I. Autiero, S. Costantini, and G. Colonna, Modeling of the bacterial mechanism of methicillin-resistance by a systems biology approach, PLoS One 4 (2009) e6226.

[40] A. Nikitin, S. Egorov, N. Daraselia, and I. Mazo. Pathway studio - the analysis and navigation of molecular networks, Bioinformatics 19 (2003) 2155–2157.

[41] P. Shannon, A. Markiel, O. Ozier, N.S. Baliga, J.T. Wang, D. Ramage, N. Amin, B. Schwikowski, and T. Ideker, Cytoscape: a software environment for integrated models of biomolecular interaction networks, Genome Research 13(2003) 2498-2504.

[42] L. Strömbäck, V. Jakoniene, H. Tan and P. Lambrix, Representing, storing and accessing molecular interaction data: a review of models and tools, Briefings in Bioinformatics 7 (2006) 331-338.

[43] M. Baitaluk et al., Pathsys: integrating molecular interaction graphs for systems biology, BMC Bioinformatics, 7 (2006) 55.

[44] K.B. Wee, U. Surana, and B.D. Aguda, Oscillations of the p53-Akt network: implications on cell survival and death, PLoS One, 4 (2009) e4407.

[45] F. Vazquez and P. Devreotes, Regulation of PTEN function as a PIP3 gatekeeper through membrane interaction, Cell Cycle, 5 (2006) 1523-1527.